\documentstyle[twoside,fleqn,espcrc2]{article}
\newcommand{\eq}{\begin{equation}}
\newcommand{\en}{\end{equation}}
\newcommand{\eqa}{\begin{eqnarray}}
\newcommand{\ena}{\end{eqnarray}}

\newcommand{\AmS}{{\protect\the\textfont2
  A\kern-.1667em\lower.5ex\hbox{M}\kern-.125emS}}

\input epsf
\title{
Disorder Parameter of Confinement
}
\author{Naoki Nakamura\address{
        Department of Physics, Kanazawa University, Kanazawa 920-11, Japan},
        Vitaly Bornyakov\address{
        Institute for High Energy Physics, 142284 Protvino, Russia
        },
        Shinji Ejiri \hspace{-2mm}\addtocounter{address}{-2}
        \addressmark\hspace{2mm},
        \addtocounter{address}{1}
        Shun-ichi Kitahara\address{
        Jumonji University, Niiza, Saitama 352, Japan
        },\\
        Yoshimi Matsubara\address{
        Nanao Junior College, Nanao, Ishikawa 926, Japan
        }
        and \hspace{2mm}
        Tsuneo Suzuki \hspace{-2mm}\addtocounter{address}{-4}
        \addressmark\hspace{2mm}
}

\begin{document}

\begin{abstract}

The disorder parameter of 
confinement-deconfinement phase transition
based on the monopole action determined previously in $SU(2)$ QCD
are investigated.
We construct an operator 
which corresponds to the order parameter
defined in the abelian Higgs model.
The operator shows proper behaviors as the disorder parameter
in the numerical simulations of finite temperature QCD.

\end{abstract}

\maketitle

\input epsf

\section{
Introduction
}

One of the challenging problems of QCD is to explain 
the confinement phenomena.
The dual superconductor scenario of confinement \cite{thooft1,mandels}
and the related idea of abelian projection \cite{thooft2} 
have obtained convincing
support in the numerical simulations of the lattice gauge theory
\cite{desy1,yotsu,suzu93}.
In this scenario, the condensation of abelian monopoles is responsible for 
the confining force. These monopoles in the non-abelian gauge theory
are defined after fixing the gauge freedom
down to the maximal abelian subgroup \cite{thooft2}.
In the compact $U(1)$ gauge theory the condensation of monopoles
in the confinement (strong coupling) phase has been
proved  in various ways \cite{poly,banks,suzuki_shiba,froehlich,wiese}. 
Also in the non-abelian gauge theory,
such picture has been shown from the viewpoint of
the energy-entropy balance \cite{suzuki_effact,kitahara}.
But the disorder parameter of confinement in the non-abelian case
has not been established yet.
The promising attempts
to prove it were made in \cite{digiacomo,polikarpov}. 
We propose a disorder parameter starting from a monopole-current
$U(1)$ action which was derived numerically
and so exactly from $SU(2)$
QCD \cite{suzuki_effact}.

The theory of monopoles in $SU(2)$ QCD
is described
by the partition function
\eqa
Z= \sum_{^*k\in Z,\delta ^*k=0}
                \exp(-S[^*k]).
\ena
where $^*k$ is an integer-valued 1-form on a dual lattice,
$S[^*k]$ is a monopole action defined 
by the following relations:
\eqa
\lefteqn{\exp(-S[^*k])}\nonumber\\
            &&=\int
           {\delta(^*k-\delta^*n)}e^{-S[U]}
              \delta(X^\pm)\Delta_F(U)DU\nonumber\\
            &&=\int{DuDc}{\delta(^*k-\delta^*n)}
              e^{ -S[u,c]}\delta(X^\pm)\Delta_F(U)\nonumber\\
            &&=\int{Du}{\delta(^*k-\delta^*n)}
                e^{ -S_{eff}[u]}.\nonumber
\ena
Here $\delta(X^{\pm})$ is the maximally abelian (MA) gauge fixing condition
and $\Delta_F(U)$ is the Faddeev-Popov ghost term.
$\delta(^*k-\delta^*n)$ gives a definition of
the monopole current.
 The abelian link variables,$u$, can be separated after 
fixing MA gauge:
\eqa
U'(s,\mu)=V(s)U(s,\mu)V^{\dag}(s+\mu)
         =c(s,\mu)u(s,\mu),\nonumber
\ena
\eqa
u(s,\mu)=\left(\begin{array}{cc}
          e^{i\theta_\mu(s)} & 0\\
          0 & e^{-i\theta_\mu(s)}
         \end{array}\right),\nonumber
\ena
\eqa
\theta_{\mu\nu}(s)=\bar
\theta_{\mu\nu}(s)+2\pi{n}_{\mu\nu}(s),
\quad (-\pi<\bar\theta_{\mu\nu}(s)\le\pi).\nonumber
\ena
$n_{\mu\nu}$ is an integer-valued plaquette variable
corresponding to Dirac string.
By extending Swendsen's method  the monopole action $S[^*k]$ 
has been determined in \cite{suzuki_effact} in the form:
\eqa
S[k_\mu]=\sum_{s,s',\mu}k_\mu(s){\cal F}(s,s')k_\mu(s').
\ena
This action is our starting point for studying
 the disorder parameter.

\renewcommand{\thefootnote}{\fnsymbol{footnote}}

\section{
Disorder parameter in $SU(2)$ QCD
}
The monopole currents action can be deformed to 
the (dual) abelian Higgs model action \cite{smit_sijs,cher_poli}
under some approximation of $S[^*k]$.
The order parameter in the abelian Higgs model on the lattice
has been defined by Kennedy and King \cite{kenn_king} in terms of
gauge invariant correlation function.
Correspondingly, we can define a disorder parameter of
confinement in $SU(2)$ QCD.

The monopole action 
determined in \cite{suzuki_effact} has been fitted as follows:
\eqa
S[^*k]=(^*k,{\cal D}^*k),\label{monac}
\ena
where
\eqa
{\cal D}=m_0(b)+\alpha(b)\tilde{\Delta}^{-1},\quad
\alpha(b)\equiv\frac12\left(\frac{4\pi}{g(b)}\right)^2,\nonumber
\ena
and $\tilde{\Delta}^{-1}$ 
is a modified Coulomb
propagator \cite{suzuki_effact,smit_sijs}.
$b$($=na$) is a renormalized lattice spacing
($a$ and $n$ imply the lattice spacing 
and the monopole extendedness, respectively)
\cite{suzuki_effact}.
$g(b)$ is the running coupling constant of $SU(2)$ QCD.
Here we assume that $\tilde{\Delta}^{-1}$ can be replaced
by the ordinary Coulomb propagator ${\Delta}^{-1}$.\footnote[2]
{More recent fitting suggests that 
$\tilde{\Delta}^{-1}$ in Eq.(\ref{monac}) is a massive propagator
rather than Coulomb one \cite{suzuki96}.
 This gives a small change in the partition function
$Z_{mon}$.
But it can be written in the $U(1)$ invariant form and we get the same
operator as the disorder parameter Eq.(\ref{dispar}).}
Then $Z_{mon}$ can be represented as
\eqa
\lefteqn{Z_{mon}=\int_{-\infty}^{\infty}D^*C
  \int_{-\pi}^{\pi}D^*{\phi}
  \sum_{^*l\in Z}}\nonumber\\
 & &{\exp}\left[-\frac1{4\alpha}
          \Vert{d}^*C\Vert^2
          -\frac{1}{4m_0}{\Vert}d^*\phi+2\pi^*l-^*C\Vert^2
          \right].\nonumber
\ena
This is the partition function for the abelian Higgs model
with non-compact dual gauge field $^*C$ and fixed length
(dual) scalar field $^*\Phi = \exp(i ^*\phi)$.
 In \cite{kenn_king} the order parameter
for the abelian Higgs model has been introduced. In our notations 
the order parameter for the abelian Higgs model \cite{kenn_king} reads:
$G^{\infty} \equiv \lim_{|x-y| \to \infty} G_{x,y}$,
\eq
G_{x,y} \equiv 
  \langle e^{i(^*\phi,\delta_x-\delta_y)} \cdot e^{-i(^*C,^*h)}\rangle.
\label{dispar}
\en
It can be rewritten in the monopole currents representation as
\eqa
\lefteqn{G_{x,y} 
=\frac{1}{Z_{mon}}
\sum_{^*k(^*c_1)\in Z,\delta ^*k=0}
    \exp\left[-m_0\Vert{^*h}\Vert^2\right]\times}\nonumber\\
& &\times\exp\left[-(^*k+^*h-^*\omega,{\cal D}({^*}k+^*h-^*\omega))
    \right],\nonumber
\ena
where ${^*\omega}$ is a string($\delta^*\omega={\delta}_x-{\delta}_y$)
 connecting $x$ and $y$,
and ${^*h}$ is a {\em smeared string} satisfying the relation:
\eqa
{\delta}{^*h}={\delta}_x-{\delta}_y.\nonumber
\ena
${^*h}$ is not unique. Following \cite{kenn_king}, we choose
\eqa
{^*h}=d\Delta^{-1}({\delta}_x-{\delta}_y).
\ena
Then $G_{x,y}$ takes the form
\eqa
\lefteqn{G_{x,y} =\frac{1}{Z_{mon}}
\sum_{^*k(^*c_1)\in Z,\delta ^*k=0}}\nonumber\\
  & & \exp[
    -(^*k-^*\omega,{\cal D}({^*}k-^*\omega))
     +\alpha
      (^*h,{\Delta^{-1}}{^*h})].\nonumber
\ena
Note that the monopole currents,${^*}k$, do not couple
to the smeared string,${^*h}$.
In this representation it seems that 
$^*h$  is not important
for the disorder parameter behavior.
But it guarantees the invariance of $G_{x,y}$
with respect to the dual $U(1)$ gauge transformation \cite{kenn_king}.

\section{Numerical simulations}

\begin{figure}
\epsfxsize=75mm\epsfysize=70mm
 \vspace{-20pt}
 \begin{center}
 \leavevmode
\epsfbox{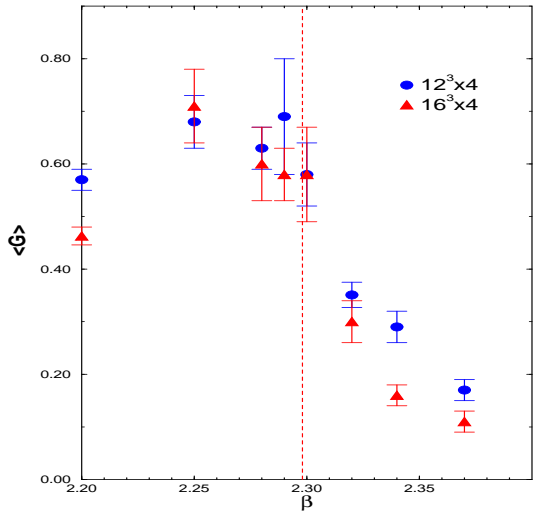}
 \end{center}
 \vspace{-37pt}
\caption{
$G_{x,y}$
 versus $\beta$ on $12^3\times 4$ and $16^3\times 4$ 
$SU(2)$ lattices.
Dashed line shows critical point.
\label{disoperator}
}
\vspace{-15pt}
\end{figure}

The simulations have been done in finite temperature $SU(2)$ QCD
on $16^3\times 4$ and $12^3\times 4$ lattices.
We have adopted $2^3$ extended monopoles
and the anti-periodic boundary conditions 
for space directions.
%
%
Figure \ref{disoperator} shows our results for $G_{x,y}$
as a function of $\beta$. 
The distance $|x-y|$ is taken to be 
6 and 8 for $12^3\times 4$ and  $16^3\times 4$ lattices respectively.
One can see clearly that $G_{x,y}$ is finite for both lattice sizes
in the confinement phase(monopole currents
condensation).
While it is small in the deconfinement phase and
it seems to go to zero in the $|x-y| \to \infty$ limit.

\section{
Conclusions
}

We have proposed 
a new approach to evaluate disorder parameter
in $SU(2)$ QCD.
Our definition shows numerically the typical behaviors
as the disorder parameter of confinement.
This suggests abelian monopoles are condensed in the confinement
phase, while they are not condensed in the deconfinement phase.
It is worth noting that our definition of the disorder parameter
corresponds to definitions used in $2d$ Ising model \cite{kadanoff}
and in $4d$ compact $U(1)$ gauge model \cite{froehlich}. 

One can take other choices for $^*{h}$.
For example we have taken also as $^*{h}$ a magnetic field,
$^*{B}=d_3\Delta_3^{-1}(\delta_x-\delta_y)\nonumber$
where $d_3$ and $\Delta_3$ are 3$d$ exterior differential
and 3$d$ laplacian which are defined on a given dual
time slice.
Then we 
obtained similar results.

In order to 
make more precise measurements of $G^{\infty}$
larger distances $|x-y|$
and consequently larger lattices are necessary.

%
It is also important that the monopole currents action $S[^*k]$ 
is determined more precisely.
This enables us to consider other effective theories
derived from the current action,
e.g. Z-gauge theory or abelian Higgs model are interesting. 
It is expected that in these models, the behaviors of
disorder parameter are clearer.

This work is financially supported by JSPS \\
Grant-in Aid for Scientific  
Research (B)\\ (No.06452028).\\
One of us (V.B.) thanks the Local
Organizing Committee and Russian Ministry of Science
for  partial support.

\end{document}